\documentclass[6pt]{article}
\def\half{\frac{1}{2}}

\newfont{\bbbold}{msbm10 scaled \magstep1}

\def\cO{\cal O}

\newfont{\goth}{eufm10 scaled \magstep1}

\def\a{\alpha}

\def\b{\beta}

\def\c{\gamma}
\def\d{\delta}
\def\e{\epsilon}\def\ve{\varepsilon}
\def\f{\phi}

\def\h{\eta}

\def\l{\lambda}\def\L{\Lambda}

\def\r{\rho}
\def\s{\sigma}\def\S{\Sigma}
\def\t{\tau}

\def\beq{\begin{equation}}\def\eeq{\end{equation}}
\def\beqa{\begin{eqnarray}}\def\eeqa{\end{eqnarray}}
\def\barr{\begin{array}}\def\earr{\end{array}}

\def\o{\omega}\def\O{\Omega}

\def\xz{\times}

\def\nab{\nabla}

\def\hT{\hat{T}}\def\hR{\hat{R}}

\def\hR{\widehat{R}}\def\hT{\widehat{T}}

\def\ghO{\widehat{\O}}

\let\la=\label

\def\nn{\nonumber}
\def\bd{\begin{document}}
\def\ed{\end{document}}
\def\ba{\begin{array}}
\def\ea{\end{array}}
\def\bea{\begin{eqnarray}}
\def\eea{\end{eqnarray}}
\def\ft#1#2{{\textstyle{{\scriptstyle #1}\over {\scriptstyle #2}}}}
\def\fft#1#2{{#1 \over #2}}
\newcommand{\be}{\begin{equation}}
\newcommand{\ee}{\end{equation}}
\newcommand{\eq}[1]{(\ref{#1})}
\newcommand{\w}[1]{\\[0.#1cm]}
\def\eqs#1#2{(\ref{#1}-\ref{#2})}
\def\det{{\rm det\,}}
\def\tr{{\rm tr}}
\newcommand{\Section}[1]{\section{#1} \setcounter{equation}{0}}
\newcommand{\hoch}[1]{$\, ^{#1}$}
\newcommand{\kings}
{\it\small Department of Mathematics, King's College, London, UK}
\newcommand{\iftsp}
{\it\small Instituto de Fisica Teorica, State University of S\~ao Paulo, Brasil}
 \makeatletter
\renewcommand\theequation{\thesection.\arabic{equation}}
\@addtoreset{equation}{section} \makeatother

\newcommand{\auth}
{\large P.S. Howe}

\begin{document}

\hfill{KCL-TH-08-04}

\vspace{30pt}

\begin{center}
{\Large{\bf Heterotic supergeometry revisited}}
 \vspace{30pt}

\auth

\vspace{15pt}

{\it\small Department of Mathematics, King's College, London, UK}

\vspace{60pt}

{\bf Abstract}

\end{center}

The superspace geometry relevant to the heterotic string is reviewed from the point of view of the off-shell supermultiplet structure of $N=1,d=10$ supergravity. The anomaly-modified seven-form Bianchi identity is analysed at order $\a'^3$ and shown to admit a complete solution. The corresponding $\a'^3$ deformation of the dimension-zero torsion tensor is derived and shown to obey the appropriate cohomological constraint.

\pagebreak \tableofcontents \setcounter{page}{1}


\section{Introduction}


Higher-order corrections to the effective field theories for the massless modes of the string beyond the leading-order supergravity theories are important for several reasons. In particular, they represent genuine stringy effects in the theory, they provide a testing ground for duality conjectures beyond the leading order and they are needed to evaluate the effects of string corrections on solutions to the supergravity equations of motion.\footnote{A recent example of the usefulness of higher-order terms is the application of $R^2$ terms in $d=5$ \cite{Hanaki:2006pj}
 to near-horizon symmetry enhancement (see \cite{Duff:2008pa} for a discussion).} For solutions which have non-vanishing flux it is necessary to know the complete bosonic actions and not just the curvature terms. The problem of obtaining the complete actions, for example at order $\a'^3$ which will be the main focus of this paper, is still unsolved even though much is known about some of the terms. There are several different approaches that have been used with partial success: the computation of string scattering amplitudes \cite{Gross:1986mw}, the calculation of beta-functions in the string sigma model \cite{Grisaru:1986kw,Candelas:1986tz,Lu:2005im}, supersymmetry for component Lagrangians \cite{Bergshoeff:1988nn,Bergshoeff:1989de,deRoo:1992zp}
and superspace methods. It is the latter we shall be concerned with in this paper. We shall start by reviewing some old results in the light of the known off-shell multiplet structure of $N=1, d=10$ supergravity; the main part of the paper is a derivation of the $\a'^3$ correction to the dimension-zero torsion due to the one-loop string correction.

The on-shell constraints for $d=10,N=1$ supergravity were first written down in \cite{Nilsson:1981bn}. The Chapline-Manton theory which includes the Chern-Simons term in the Yang-Mills sector was described in \cite{Kallosh:1985cd,Nilsson:1986md,Atick:1985de}. This theory is a consistent approximation at order $\a'$, but the inclusion of the Lorentz Chern-Simons term is more difficult since it induces corrections at all orders. It has been studied in detail in references \cite{D'Auria:1987yv,Raciti:1989je,Bonora:1986ix,Bonora:1987xn,Bonora:1990mt,Bonora:1992tx}
(see \cite{Lechner:2008uz} for a recent update), from a slightly different viewpoint in \cite{Bellucci:1988ff,Bellucci:1990fa,Gates:2004cd}, and from the dual perspective in
\cite{Saulina:1995eq,Saulina:1996vn}. In this paper we shall start from the off-shell version of $N=1,d=10$ supergravity given in reference \cite{Howe:1982mt}. Up to order $\a'^2$ this can be simplified to a $128+128$ multiplet which was described in superspace in \cite{Nilsson:1985si,Howe:1986ed}. We shall show how this multiplet fits into the approach of the Italian school in the next section. At order $\a'^3$, the off-shell structure would seem to indicate that the dimension-zero torsion tensor should be deformed \cite{Nilsson:1986cz,Howe:1986ed}. This is also required by the fact that the Bianchi identity for the seven-form must be modified at this order, as noted in \cite{Candiello:1994ew}.

The organisation of the paper is as follows: in the next section the off-shell structure of the theory is reviewed and applied to supergravity in the presence of the Lorentz Chern-Simons term for the three-form up to order $\a'^2$; in section 3 we recap some cohomological results and describe how they can be applied to superspace deformation theory; in section 4, the main section, we use these methods to show that a complete solution to the modified seven-form Bianchi identity can be found at order $\a'^3$,  we explicitly give the $\a'^3$ deformation of the dimension-zero torsion and we show that this deformation does indeed obey the correct cohomological constraint. We conclude with some remarks about the inclusion of the Yang-Mills fields and some speculations about the string tree level $\a'^3$ term. The paper can be viewed as a companion to \cite{Berkovits:2008qw}
where the problem of higher-order corrections in the  heterotic theory was investigated from the point of view of integral invariants constructed using cohomology and the super-form method.


\section{The theory at $\a'^2$}



\subsection{The multiplet structure of $N=1,d=10$ supergravity}


The off-shell structure of supergravity was derived in \cite{Howe:1982mt} starting from the supercurrent for $N=1,d=10$ super Yang-Mills (SYM) theory. In ten dimensions it turns out that it is not enough to use the free SYM theory for this purpose because it contains a number of conserved currents in unusual representations of the Lorentz group that are not conserved in the interacting theory \cite{Bergshoeff:1981um}. The full supercurrent has two parts: a $128+128$ multiplet comprised of a conserved traceless energy-momentum tensor, an identically conserved six-form current which couples to the six-form gauge field in the dual version of $N=1,d=10$ supergravity and a gamma-traceless vector-spinor, and an entire scalar superfield. Ten-dimensional Yang-Mills theory is, of course, not superconformal, so that the separation of the supercurrent into the above two parts cannot be carried out locally. The $128+128$ multiplet was written down first in \cite{Bergshoeff:1982av} where its non-local character was also discussed. If we denote the SYM field strength superfield by $\L^\a$ the supercurrent is

\be
 J_{abc}=Tr (\L \c_{abc} \L)\ .
 \la{2.1}
\ee

The supergravity multiplet dual to the supercurrent has a similar structure. The $128+128$ multiplet consists of the graviton, the gravitino and the six-form field strength, together with constraints on the curvature scalar and the double-gamma-trace of the gravitino field strength. This multiplet is local, and can be thought of as being partially off-shell. In order to go completely off-shell one has to introduce an entire scalar superfield dual to the one in the supercurrent. It turns out to have dimension minus six, and so cannot be non-zero on-shell until order $\a'^3$. When this multiplet is not put on-shell the constraints on the graviton and gravitino field strengths are no longer present and the theory is completely off-shell. It is clear that one cannot write down a Lagrangian with these fields; in order to do so it is necessary to introduce a new unconstrained scalar superfield which has dimension zero. Its leading components are the dilaton and the dilatino, thus completing the physical fields of the supergravity multiplet. In the on-shell theory this superfield becomes the supergravity field strength superfield whose independent components are the dilaton, the dilatino and the field strengths of the two-form potential, the gravitino and the graviton.


\subsection{The $128+128$ geometry}


It is straightforward to write down the superspace constraints corresponding to the partially off-shell $128+128$ multiplet. The non-vanishing components of the torsion may be chosen to be \cite{Howe:1986ed}

\bea
 T_{\a\b}{}^c&=&-i(\c^c)_{\a\b}  \nn\w1
 T_{a\b}{}^\c&=& (\c^{bc})_\b{}^\c G_{abc} +\frac{1}{6} (\c_{abcd})_\b{}^\c G^{bcd}\nn\w1
 T_{ab}{}^\c&=& \Psi_{ab}{}^\c\ .
 \la{2.2}
\eea

The field $G_{abc}$ is the dual of the seven-form field strength, while $\Psi_{ab}$ is the gravitino field-strength. The components of the curvature tensor are

\bea
 R_{\a\b,cd}&=&4i( (\c^e)_{\a\b} G_{cde} +\frac{1}{12} (\c_{cdefg})_{\a\b} G^{efg})\nn\w1
 R_{\a b,cd}&=&\frac{i}{2}(\c_b \Psi_{cd}-\c_d\Psi_{bc}-\c_c\Psi_{db})_\a\ ,
 \la{2.3}
\eea

as well as the usual curvature tensor $R_{ab,cd}$ at dimension two. The non-vanishing components of the seven-form field strength are

\bea
 H_{abcde\a\b}&=&-i(\c_{abcde})_{\a\b}\nn\w1
 H_{abcdefg}&=&-2\ve_{abcdefghij} G^{hij}\ ,
 \la{2.4}
\eea

the first of which may be written as $H_{5,2}=-i\c_{5,2}$. Here, a $(p,q)$-form is one with $p$ even and $q$ odd indices, as usual, and $\c_{p,2}$ denotes a symmetric $p$-index gamma-matrix considered as a $(p,2)$-form. The supersymmetry transformations of the field strengths are given by the differential identities

\bea
 \nab_\a G_{abc}&=&-\frac{i}{24} (\c_{abcde} \Psi^{de})_\a\nn\w1
 \nab_\c\Psi_{ab}{}^\d&=&R_{ab,\c}{}^\d -2 \nab_{[a} T_{b]\c}{}^\d -([T_a,T_b])_\c{}^\d\ ,
 \la{2.5}
\eea

where $T_a$ denotes the dimension-one torsion considered as a matrix-valued covector. In addition,

\bea
 \c^{ab}\Psi_{ab}&=&0 \nn\w1
 R&=&12 G^{abc} G_{abc}
 \la{2.6}
\eea

are the non-linear constraints on the partially off-shell field strengths.

It is worth noting that this multiplet is determined by the geometrical constraints alone; in other words, given the torsion constraints of \eq{2.2} one can construct a closed seven-form with components given in \eq{2.4}.

The geometrical constraints given here form the basis of the Italian school's approach to heterotic supergeometry. Indeed, it has been argued \cite{Bonora:1990mt} that it is only necessary to impose the standard dimension-zero torsion constraint and that the rest follow from this. On the other hand, if one works in Weyl superspace, i.e. if one includes an abelian scale connection, then the standard dimension-zero constraint does not imply that the scale curvature vanishes \cite{Bonora:1990mt}, as it does in the $d=11$ case \cite{Howe:1997rf}. Starting from this Weyl geometry and reducing back to a Lorentzian structure group would therefore seem to imply the presence of additional fields. Whether or not this is the case is not relevant to the theory up to order $\a'^2$ for which we need to adopt the above constraints corresponding to the $128+128$ partially on-shell multiplet. On the other hand, at $\a'^3$, we shall see that there is an additional spinor field, not directly determined by the modified dimension-zero torsion, in the dimension one-half torsion.


\subsection{Equations of motion}


A convenient way of obtaining the equations of motion, used by the Italian school, is to introduce the three-form field strength $H_3$ dual to the seven-form $H_7$. Its Bianchi identity is

\be
 dH_3=k_3 X_4
 \la{2.7}
\ee

where $k_3\propto \a'$, and $X_4=Tr R^2$ in the gravitational sector. The key BPT theorem \cite{Bonora:1986ix} states that, given the above geometrical constraints, $X_4$ is exact up to an additional closed four-form that vanishes in the $(0,4)$ and $(1,3)$ sectors.  Thus we have

\be
 X_4= d Q_3 + S_4\,;\qquad {\rm where}\qquad \ S_{4,0}=S_{3,1}=0\ .
 \la{2.8}
\ee

This theorem is neither obvious nor easy to prove. With its aid it becomes much easier to solve \eq{2.7} by taking $Q_3$ over to the left-hand side. We shall not go into the details here since they can be found in the literature, see \cite{Lechner:2008uz} for a recent discussion, but merely give the leading-order terms in $H_3$,

\bea
 H_{0,3}&=&0 \nn\w1
 H_{1,2}&=&-iS \c_{1,2} + \cO (\a')\nn\w1
 H_{ab\c}&=&-(\c_{ab} \l)_\c + \cO (\a')\nn\w1
 H_{abc}&=& 12S G_{abc}+ \cO (\a')\ .
 \la{2.9}
\eea

Here $S$ is related to the dilaton $S:=\exp(-2\f/3)$ and $\l_\a:=\nab_\a S$ is the dilatino. At this stage one sees that there are only physical fields remaining and that they must therefore be on-shell. Note that the frame we are using here is the so-called brane-frame, the metric in this frame being related to that of the string frame by

\be
 g_B=\exp(-2\f/3)g_S\ .
 \la{2.10}
\ee

In summary, the heterotic theory up to $\a'^2$ is described by the torsion constraints given above together with the three-form gauge-field strength tensor just introduced. At order $\a'^3$ the off-shell structure of the supergravity theory suggests that the dimension-zero torsion should be deformed; in any case, such a deformation is needed to incorporate the one-loop string anomaly term.

We conclude this section with a brief comment on the work of references \cite{Bellucci:2006cx,O'Reilly:2006px} where it has been argued that a solution to the heterotic Bianchi identities needs a correction to the dimension-zero torsion at $\a'^2$. As we have seen, the off-shell structure of supergravity does not suggest that this is the case, but this does not mean that these papers are incorrect; it could be that the formalism of \cite{Bellucci:2006cx,O'Reilly:2006px} is related to the formalism advocated here by suitable (and presumably rather complicated) field redefinitions.


\section{Deformations and cohomology}


For theories with maximal supersymmetry, and therefore no (known) auxiliary fields, a systematic way of determining higher-order corrections is via deformation theory in superspace. This involves spinorial cohomology \cite{Cederwall:2001dx,Howe:2003cy} which in it simplest form is equivalent to pure spinor cohomology \cite{Howe:1991mf,Howe:1991bx,Berkovits:2000fe,Berkovits:2002zk}. In this section we shall briefly review this formalism following, for the most part, reference \cite{Berkovits:2008qw}. We denote by $\O_{p,q}$ the space of $(p,q)$-forms and write the exterior derivative as

\be
d=d_0 + d_1 + t_0 + t_1,
\la{3.1}
\ee

with bi-degrees
$(1,0),(0,1),(-1,2),(2,-1)$ respectively \cite{Bonora:1986ix}. It is easiest to write these using covariant derivatives
and the torsion. Thus $d_0\sim E^a(\nab_a + T_{a\cdot}{}^{\cdot})$ and $d_1\sim E^\a(\nab_\a +
T_{\a\cdot}{}^{\cdot})$ are even and odd derivatives while $t_0$ (see \eq{3.3}) and $t_1$ are algebraic operations
involving the dimension zero and three-halves components of the torsion tensor.

The identity $d^2=0$, when decomposed into bi-degrees, includes the following components,

\bea
 t_0^2 &=& 0\nn\w1
 d_1 t_0 + t_0 d_1 &=& 0\nn\w1
 d_1^2 + d_0 t_0 + t_0 d_0&=& 0\ .
 \la{3.2}
\eea

The first of the above
equations allows us to introduce the cohomology groups $H_t^{p,q}$, the space of $t_0$-closed
$(p,q)$-forms modulo the $t_0$ exact ones \cite{Bonora:1986ix}. The groups $H_t^{0,q}:=H_t^q$ can be thought of as spaces of (generalised)
pure multi-spinors.

We can also define $t_0$-cohomology groups for $(0,q)$-forms taking their values in $\wedge^k T_0$; these will turn out to be useful for finding the $H_t^{p,q}$ groups. To do this let us first define the space $\O_{p,q}^{k,l}$ consisting of $(p,q)$ forms taking their values in $\wedge^k T_0\otimes \wedge^l T_1$, i.e the space of $(p,q)$-forms which are also $(k,l)$-multivectors. The dimension-zero torsion can be made to act in two ways on this space: firstly, we define $t_0$ to act as before, i.e. ignoring the multivector indices, and secondly we define a new operation $t^0:\O_{p,q}^{k,l}\rightarrow\O_{p,q+1}^{k+1,l-1}$. In components these operations are given by

\be
 (t_0 \o)_{a_1\ldots a_{p-1},\a_1\ldots \a_{q+2}}^{b_1\ldots b_k,\b_1\ldots\b_l}=\frac{(q+1)(q+2)}{2} T_{(\a_1\a_2}{}^c \o_{|c a_1\ldots a_{p-1}|,\a_{3}\ldots \a_{q+2})}^{b_1\ldots b_k,\b_1\ldots\b_l}\ ,
 \la{3.3}
\ee

and

\be
 (t^0\o)_{a_1\ldots a_p,\a_1\ldots\a_{q+1}}^{b_1\ldots b_{k+1},\b_1\ldots\b_{l-1}}=
 (-1)^{p+q+1}(k+1)(q+1)\o_{a_1\ldots a_p,(\a_1\ldots\a_q}^{[b_1\ldots b_k,|\b_1\ldots\b_{l-1}\c|}T_{\a_{q+1})\c}{}^{b_{k+1}]}\ .
 \la{3.4}
\ee

It is straightforward to show that $t:=t_0+t^0$ is nilpotent,

\be
 t^2=0\qquad \Leftrightarrow\qquad (t_0)^2=(t^0)^2=t_0 t^0+t^0t_0=0 \ .
 \la{3.5}
\ee

The operation $t$ maps $\oplus\, \O_{p-r,q+r}^{k-r,l+r}$ to $\oplus\, \O_{p-r-1,q+r+2}^{k-r,l+r}$ where the sum is over all integers $r$. We shall be interested in the cohomology groups $(H_t)^{k,0}_{0,q}:=H_t^q(\wedge^k T_0)$. Since elements of $\O_{0,q}^{k,0}$ are annihilated by $t_0$ and $t^0$, this group is given by elements of this space modulo elements of the form $t_0 \l+t^0\r$ where $\l\in \O_{1,q-2}^{k,0}$ and $\r\in\O_{0,q-1}^{k-1,1}$.

The groups $H_t^{p,q}$ will form the starting point for the analysis of the cohomology groups we
are interested in. The non-vanishing $H_t^{p,q}$ cohomology groups for $N=1,d=10$ are\footnote{This is based on the assumption that the cohomology is generated by $\c_{1,2}$ and $\c_{5,2}$.}

\bea
 H_t^{0,q}&=& H_t^q \nn\w1
 H_t^{1,1}&=&\O_{0,0}^{1,0}\nn\w1
 H_t^{1,q}&=&H_t^{q-2}(\L^4 T_0) + \d_{q 2}\,\O_{0,0}^{0,0},\ q\geq 2\nn\w1
 H_t^{p,q}&=&H_t^{q-2}(\L^{5-p}T_0),\ q\geq 2;\ p\in \{2,3,4,5\}\ .
\la{3.5.1}
\eea

We define an odd derivative $d_s$ which acts on elements of $H_t^{p,q}$ by

\be
 d_s [\o] :=[d_1 \o] \ ,
 \la{3.6}
\ee

where the square brackets denote equivalence classes in $H_t$ \cite{Howe:2003cy}. With the aid of \eq{3.2} it is easy to check that $d_s$ is well-defined and squares to zero so that we can define the spinorial cohomology groups $H_s^{p,q}$
in the obvious way: $H_s^{p,q}:= H_{d_s}(H_t^{p,q})$, with $H_s^{0,q}:=H_s^q$.

We shall also need spinorial cohomology groups for $(0,q)$-forms taking their values in $\wedge^k T_0$. Let $h\in\O_{0,q}^{k,o}=\O_{0,q}(\wedge^k T_0)$. We can define an odd exterior derivative on such objects as follows:

\bea
 (d_1 h)_{\a_1\ldots \a_{q+1}}^{a_1\ldots a_k}&=&(q+1) \nab_{(\a_1} h^{a_1\ldots a_k}_{\a_2\ldots \a_{q+1})}+\frac{q(q+1)}{2}T_{(\a_1\a_2}{}^\c h_{|\c|\a_3\ldots\a_{q+1})}^{a_1\ldots a_k} \nn\w2
 &\phantom{=}& +\, (-1)^{q+1}k(q+1)h_{(\a_1\ldots\a_q}^{[a_1\ldots a_{k-1}|b|}T_{\a_{q+1})b}{}^{a_k]}\ .
 \la{3.8}
\eea

A straightforward computation shows, provided the dimension zero torsion is covariantly constant, that

\be
 d_1^2 h=t_0 \l + t^0 \r
 \la{3.9}
\ee

for some (computable) $\l\in\O_{1,q}^{k,0}$ and $\r\in\O_{0,q+1}^{k-1,1}$. We can therefore define $d_s[h]=[d_1 h]$ and $H_s^q(\wedge^k T_0)=H_{d_s}(H_t^q(\wedge^k T_0))$.

We now briefly review how one can apply this formalism to deformation theory. A simple example is provided by $N=1,d=10$ SYM in flat superspace \cite{Cederwall:2001bt,Cederwall:2001td}. The curvature two-form is $F=dA+A^2$ where $A$ is the potential. At lowest order the field equations are determined by the constraint $[F_{0,2}]=0$. This means that we can define a covariant spinorial derivative $D_s$ acting on Lie-algebra valued fields which squares to zero and hence defines a cohomology theory in exactly the same way as described above. The Bianchi identity for the first-order correction, $\stackrel{(1)}{F_{0,2}}$, is then equivalent to

\be
 D_s [\stackrel{(1)}{F_{0,2}}]=0\ ,
 \la{3.10}
\ee

i.e. the first-order deformation is given by an element of $H_s^2$, where the coefficients have to be tensorial functions of the physical fields. We are using cohomology because an exact deformation can be removed by a field redefinition of the potential. At the next order we find

\be
 D_s [\stackrel{(2)}{F_{0,2}}]+\chi_{0,3}=0\
 \la{3.11}
\ee

where $\chi_{0,3}$ is obtained by including the effect of the first-order deformation on the equations of motion to the spinorial derivative of the first-order deformation of $F_{0,2}$. This is a function which can be computed explicitly in terms of the physical fields and which must clearly be exact in order for \eq{3.11} to have a solution, i.e. $\chi_{0,3}$ must be trivial in $H_s^3$. If this is the case, then any further deformation at this order will again be given by an element of $H_s^2$, although it will have a different dimension as the expansion parameter is related to $\a'$. The key point here is that, provided there are no higher cohomological obstructions, any new possibilities that can arise are always determined by the cohomology associated with the zeroth-order theory.

Now let us consider supergravity \cite{Cederwall:2001dx,Howe:2003cy}. The above technique will turn out to be useful for $N=1,d=10$ even though auxiliary fields exist. We shall be interested in the constraint equation which is obeyed by the deformed dimension-zero torsion. It turns out that the possible deformations are given by $H_s^2(T_0)$, computed with respect to the lowest-order theory. To see this consider the dimension one-half Bianchi identity

\be
 \nab_{(\a} T_{\b\c)}{}^d + T_{(\a\b}{}^e T_{e\c)}{}^d + T_{(\a\b}{}^\e T_{|\e|\c)}{}^d=0\ .
 \la{3.12}
\ee

At order $\a'^3$, this equation reads

\be
 \nab_{(\a} \stackrel{(3)}{T}_{\b\c)}{}^d + \stackrel{(0)}{T}_{(\a\b}{}^e \stackrel{(3)}{T}_{e\c)}{}^d + \stackrel{(0)}{T}_{(\a\b}{}^\e \stackrel{(3)}{T}_{|\e|\c)}{}^d=-\stackrel{(3)}{T}_{(\a\b}{}^e \stackrel{(0)}{T}_{e\c)}{}^d -\stackrel{(3)}{T}_{(\a\b}{}^\e \stackrel{(0)}{T}_{|\e|\c)}{}^d\ ,
 \la{3.13}
\ee

where $\stackrel{(0)}{T}$ and $\stackrel{(3)}{T}$ denote the zeroth- and third-order torsion components. Looking at the definitions given above we see that this equation is of the form of \eq{3.9} where $t_0$ and $d_1$ are defined with respect to the lowest-order theory and where $h$ is the $\a'^3$ contribution to the dimension-zero torsion, i.e. we can rewrite it as

\be
 d_s[\stackrel{(3)}{T^1{}_{,2}}]=0\ ,
 \la{3.14}
\ee

where $T^1{}_{,2}$ denotes the dimension-zero torsion considered as a $T_0$-valued $(0,2)$-form. Note that \eq{3.13} is correct at this order since there are no $\a'$ or $\a'^2$ corrections to $T_{\a\b}{}^c$.


\section{The theory at $\a'^3$}


In this section we shall compute the $\a'^3$ correction to the dimension-zero torsion using the modified Bianchi identity for the seven-form field strength $H_7$,

\be
 dH_7=k_7 X_8
 \la{4.1}
\ee

where $k_7\propto \a'^3$ and $X_8$ is an invariant eight-form constructed from the curvature. There are two of these, $Tr R^4$ and $(Tr R^2)^2$. Any linear combination turns out to be allowed by supersymmetry at this order but the relative coefficient is fixed by the Green-Schwarz anomaly-cancelling mechanism \cite{Green:1984sg} (see \cite{Candiello:1994ew} for a superspace discussion). Due to the presence of the $k_7$ factor we can compute the right-hand side at zeroth order in $\a'$. To do this it is convenient to make a field-redefinition.


\subsection{Field redefinition}


It was shown in \cite{Bonora:1986ix} that it is possible to redefine the connection in such a way that $R_{1,1}$ vanishes at zeroth order. We set $\O'=\O+\S$, where

\bea
 \S_{\a,bc}&=&-\half S^{-1}(\c_{bc}\l)_\a\nn\w1
 \S_{a,bc}&=&6G_{abc}-24i S^{-2} l_{abc}-S^{-1}\h_{a[b}\nab_{b]}S\ ,
 \la{4.2}
\eea

with $l_{abc}:=\l\c_{abc}\l$. It is then straightforward to verify that $R'_{\a\b,cd}=0$. Using the $DR=0$ Bianchi we find

\be
 R'_{\a b,cd}=(\c_b \L_{cd})_\a\ ,
 \la{4.3}
\ee

where

\bea
 \L_{ab}&=&i\psi_{ab}+S^{-1}(-\frac{3}{4}\c_{[a}\nab_{b]}\l-G_{abc}\c^c\l+2 G_{[a}{}^{cd}\c_{b]cd}\l+\frac{1}{6}G^{cde}\c_{abcde}\l)\nn\w1
 &\phantom{=}&-\frac{1}{4}S^{-2}(4\nab_{[a}S\c_{b]}\l+ \nab^c S\c_{abc}\l)
 +24iS^{-3}\l^3_{ab}\ .
 \la{4.4}
\eea

Here we have used the decomposition of the gravitino field-strength into its irreducible parts, $\Psi_{ab}=\psi_{ab}+\c_{[a}\psi_{b]}$, and $\l^3_{ab}$, which is also gamma-traceless, is defined by $l_{abc}\l=\c_{[a}\l^3_{bc]}$. To derive \eq{4.4} one has to make use of the gravitino and dilatino equations of motion,

\bea
 \psi_a&=&\frac{iS^{-1}}{4}\left(G_{abc} \c^{bc}\l-\frac{1}{6}G^{bcd} \c_{abcd}\l -\nab_a\l\right)\ ,\nn\w1
 \c^a\nab_a\l&=&-\frac{1}{6}G^{abc} \c_{abc}\l\ .
 \la{4.5}
\eea

 There are also changes to the torsion tensor which are easy to compute; in particular, the dimension one-half torsions are no longer zero. For the rest of this section we shall work with this form of the geometry but we shall drop the primes.


\subsection{Components of $X_8$}


 Because $R_{1,1}=0$ the lowest-order non-vanishing component of $X_8$ is $X_{4,4}$. Since $X_8$ is closed it follows that $t_0 X_{4,4}=0$ (where here and throughout this section $t_0$ refers to the zeroth-order torsion). From the known structure of the $H_t$ groups \eq{3.5.1} it follows that

\be
 X_{4,4}=-i\c_{5,2} Y^1{}_{,2}+ t_0 Y_{5,2}\ ,
 \la{4.5.1}\
\ee

where $Y^1{}_{,2}$ is a $T_0$-valued $(0,2)$-form, and where its vector index is contracted with one of the even indices of $\c_{5,2}$. We are going to identify $k_7 Y^1{}_{,2}$ with the $\a'^3$ correction to the dimension-zero torsion. Before solving \eq{4.1} we shall show that $Y^1{}_{,2}$ does indeed satisfy the right equation for the torsion deformation. The next order of $dX_8=0$ reads

\be
 d_1 X_{4,4} + t_0 X_{5,3}=0 .
 \la{4.6}
\ee

Substituting the expression for $X_{4,4}$ given in \eq{4.5.1} into \eq{4.6} we find

\be
-i \c_{5,2} d_1 Y^1{}_{,2} + t_0(X_{5,3}-d_1 Y_{5,2})=0\ ,
 \la{4.7}
\ee

where $d_1$ is the correct derivative acting on $T_0$-valued $(0,2)$-forms. Since \eq{4.7} is true, because $X_8$ is constructed from a curvature two-form which satisfies the standard Bianchi identity, it follows that the first term on the left must be $t_0$-exact. To see what this implies we write

\be
 (d_1 Y^1{}_{,2})^a_{\a\b\c}=(\c^b)_{(\a\b} \chi_{\c) b}{}^a+ (\c^{bcdef})_{(\a\b}\chi_{\c)abcde}{}^f\ .
\la{4.8}
\ee

The first term on the right is easily seen to give an exact contribution to \eq{4.7}, so we only need to worry about the second term. Consider the following expression,

\be
 \hat\chi_{abcde,f}=\chi_{abcde,f} +\frac{20}{14}\c_{[ab}\r_{cde]f} -\frac{15}{14} \c_f \c_{[a}\r_{bcde]}\ ,
 \la{4.9}
\ee

where

\be
 \r_{abcd}:=\chi_{abcde,}{}^e
 \la{4.10}
\ee

is easily seen to be gamma-traceless and thus to transform under the irreducible $90.16$-dimensional representation of the spin group. It is not difficult to show that $\hat\chi$ is completely gamma-traceless, traceless, anti-self-dual and that its totally antisymmetric part vanishes. It transforms under the irreducible $330.16$-dimensional representation. The contribution made by $\hat\chi$ to the first term in \eq{4.7} has the form

\be
 \c_{abcde}\xz \c_{ijklm}\xz \hat\chi^{ijklm,e}\ ,
 \la{4.11}
\ee

where the $\xz$ symbol denotes total symmetrisation over the spinor indices. This expression is clearly not exact. One can explicitly check this by contracting its spinor indices with five factors of a pure spinor $u$. The quantity $u\c_{abcde}u$ is a totally null self-dual five-form and the product of two such objects is automatically in the irreducible representation given by a Young tableau with two columns of five boxes; multiplying this by $\hat\chi$ then gives a non-vanishing four-form. Since $u\c^a u=0$, this object would have had to have been zero if the expression in \eq{4.11} were to have been exact. The fact that this is not identically true means that $\hat\chi_{abcde,f}$ must itself be zero, so that the contribution from $\chi_{abcde,f}$ in \eq{4.8} is given solely by $\r_{abcd}$. This tensor must enter into $d_1 Y^1{}_{,2}$ in the form given in the right-hand side of \eq{4.9}. It is easy to show that only the term of the form $\c_f\c_{[a}\r_{bcde]}$ survives in $d_1 Y^1{}_{,2}$. Putting all this together we find that

\be
 d_1 Y^1{}_{,2}=t_0 Z^{1,0}_{1,1} + t^0 Z^{0,1}_{0,2}\ ,
 \la{4.12}
\ee

where $Z^{1,0}_{1,1}$  is proportional to the $\chi_{\c b}{}^a$ contribution to $d_1 Y^1{}_{,2}$, and where $Z^{0,1}_{0,2}$ contains $\r$,

\be
 Z_{\a\b}{}^\c\sim (\c^{abcde})_{\a\b}(\c_a\r_{bcde})^\c\ .
 \la{4.13}
\ee

We therefore see that this will lead to a solution of the deformation equation \eq{3.13} if we take $\stackrel{(3)}{T_{\a\b}{}^c},\ \stackrel{(3)}{T_{\a\b}{}^\c}$ and $\stackrel{(3)}{T_{\a b}{}^c}$ to be proportional to $k_7$ times $Y^1{}_{,2},\ Z^{0,1}_{0,2}$ and $Z^{1,0}_{1,1}$ respectively.


\subsection{The seven-form Bianchi identity}


We now turn our attention to \eq{4.1} which we rewrite as

\be
 I_8:=dH_7-k_7 X_8= 0\ .
 \la{4.14}
\ee

Let assume that all components of $H$ up to and including $H_{4,3}$ are zero. This is certainly compatible with the structure of $X_8$. The $(4,4)$ component of $I_8$ allows us to solve for $H_{5,2}$ and $\stackrel{(3)}{T^1{}_{,2}}$ (using form notation for the dimension-zero torsion). Therefore we can assume that $I_{4,4}=0$, i.e. that this equation has been solved consistently. Since $dI=0$ identically we then have

\be
 t_0 I_{5,3}=0 \Rightarrow I_{5,3}=\c_{5,2} J_{0,1} + t_0 J_{6,1}\ .
 \la{4.15}
\ee

The significance of this is that we only have to verify the two parts of $I_{5,3}=0$ corresponding to the $J$s. Since $(dH_7)_{5,3}$ has a term $t_0 \stackrel{(3)}{H_{6,1}}$, it follows that $J_{6,1}=0$ merely enables us to solve for
$\stackrel{(3)}{H_{6,1}}$ and so cannot lead to a problem. Let as assume for a moment that $J_{0,1}=0$ does not cause a problem either; if this is the case, we have $I_{5,3}=0$ so that $t_0 I_{6,2}=0\Rightarrow I_{6,2}=t_0 J_{7,0}$. Now we can use $J_{7,0}$ to solve for $\stackrel{(3)}{H_{7,0}}$ and go on to conclude that the entire Bianchi identity is satisfied as $I_{6,2}=0\Rightarrow I_{7,1}=I_{8,0}=0$. So the only possible obstruction to a complete solution is given by $J_{0,1}$ in $I_{5,3}$. Such a term can arise from a $t_0$-closed but not exact term in $X_{5,3}$ which will have the same structure, i.e. $\c_{5,2} \o_{0,1}$, and which does not vanish as can be seen by inspection. Clearly this term does not take part in \eq{4.6} so that $\o$ will not appear in the dimension one-half torsions that we have determined by solving \eq{4.12}. Fortunately, it turns out that if one takes

\be
 T_{\a\b}{}^\c= \d_{(\a}{}^\c \t_{\b)}\qquad {\rm and} \qquad T_{\a b}{}^c=\t_{\a} \d_b{}^c\
 \la{4.16}
\ee

the field $\t$ drops out of the dimension one-half torsion Bianchi identity but contributes in just the right way via $\stackrel{(3)}{T} \stackrel{(0)}{H}$ terms to $(dH_7)_{5,3}$; in other words, $J_{0,1}=0$ can be used to solve for $\t$ in terms of $\o_{0,1}$. We therefore conclude that there is a complete solution to the $dH_7$ Bianchi identity \eq{4.1} given the starting assumption of $H$ being trivial up to order $H_{4,3}$.


\subsection{Details of $I_{4,4}$}


Writing out $I_{4,4}=0$ explicitly we have

\bea
  k_7 X_{abcd\a\b\c\d}&=&-6i k_7(\c_{abcde})_{(\a\b} Y^e_{\c\d)}+(\c^e)_{(\a\b} Y_{abcde\c\d)})\nn\w1
  &=&6T_{(\a\b}{}^e H_{abcde \c\d)}\ ,
  \la{4.17}
\eea

from which it follows, on splitting $T^1{}_{,2}$ and $H_{5,2}$ into their zeroth- and third-order terms, that

\bea
 \stackrel{(3)}{T^1{}_{,2}}&=&  k_7 Y^1{}_{,2}\nn\w1
 \stackrel{(3)}{H_{5,2}}&=& k_7 Y_{5,2}\ .
 \la{4.18}
\eea

It remains to evaluate $X_{4,4}$ explicitly. We have

\bea
 Tr(R_{1,1})^4&=&E^{\d\c\b\a} E^{dcba} (\c_a)_{\a\a'}(\c_b)_{\b\b'}(\c_c)_{\c\c'}(\c_d)_{\d\d'}
 Tr(\L^{\a'}\L^{\b'}\L^{\c'}\L^{\d'})\nn\w1
 &:=&\frac{1}{16.4!4!}E^{\d\c\b\a} E^{dcba}(\c_a \c^{ijk}\c_b)_{\a\b} (\c_c\c^{lmn}\c_d)_{\c\d} \bar M_{ijk,lmn}\ ,
 \la{4.19}
\eea

where the second line defines the tensor $\bar M$. $(Tr (R_{1,1})^2)^2$ has exactly the same form as the second line of \eq{4.19} with

\be
 \bar M_{ijk,lmn}=L_{ijk} L_{lmn}\ ,
 \la{4.20}
\ee

where

\be
 L_{ijk}:=Tr(\L \c_{ijk} \L)\ .
 \la{4.21}
\ee

In both case we can set

\be
 \bar M_{ijk,lmn}=M_{ijk,lmn} + M_{ijk[lm,n]}
 \la{4.22}
\ee

in terms of symmetry types; i.e. the first term corresponds to a Young tableau with two columns of three boxes, while $M_{ijklm,n}$ corresponds to a tableau with one column of five boxes and one of one.  For $(TrR^2)^2$, all possible traces can in principle be present, while for $TrR^4$ the second term is in the anti-self-dual $1050$ representation while the first has only the $770$ which has a single trace,

\be
 M_{ijk,}{}^{lmn}(770)=\d_{[i}{}^{[k} N_{jk],}{}^{lm]}\ ,
 \la{4.22.1}
\ee

where $N_{ij,kl}$ has the symmetries of the Weyl curvature tensor.

In indices, we therefore have

\be
 16 X_{abcd\a\b\c\d}=\left(\c_{[ab}{}^{ijk}\xz \c_{cd]}{}^{lmn} + 12 \d_{[ab}^{ij}\c^k\xz \c_{cd]}{}^{lmn}+ 36 \d_{[ab}^{ij}\c^k \xz \d_{cd]}^{lm} \c^n\right)_{\a\b\c\d}\,\bar M_{ijk,lmn}\ .
 \la{4.23}
\ee

It is not difficult to verify that the $\c_1 \xz \c_1$ and $\c_1\xz \c_5$ terms are exact and therefore only contribute to $H_{5,2}$.  With a bit more work one can show that the $\c_5 \xz \c_5$ term multiplying $M_{ijk,lmn}$ is also exact. The final contribution is

\be
 \c_{[ab}{}^{ijk}\xz \c_{cd]}{}^{lmn} M_{ijklm,n}=\frac{1}{10} \c_{abcde}\xz \c_{ijklm} M^{ijklm,e} +\ {\rm exact}
 \la{4.24}
\ee

The trace part of $M_{ijklm,n}$, which is present in $(Tr R^2)^2$, also reduces to an exact contribution.

Putting this altogether we finally arrive at

\be
 \stackrel{(3)}{T_{\a\b}{}^f}=\frac{i}{6.10.16}  k_7 (\c^{abcde})_{\a\b}\hat M_{abcde,}{}^f
 \la{4.25}
\ee

where the hat indicates that only the $1050^{-}$ representation is present.


\section{Discussion}


In this paper we have seen how the off-shell structure of $N=1,d=10$ supergravity illuminates the geometry of the effective field theory for the heterotic string and how to accommodate the one-loop $R^4$ term by modifying the dimension-zero torsion at order $\a'^3$. We note that it is straightforward to amend these results to include the Yang-Mills sector. Since we only need the zeroth-order field-strength in $X_8$, and since $F_{0,2}=0$ at this order, it follows that the entire analysis of the $H_7$ Bianchi identity will only be modified by additional contribution to the $M$-tensors, which will, moreover, have the same representational content. It is likewise straightforward to include $F$ in the $H_3$ Bianchi, although one will have to include the $\a'^2$ deformation of $F_{0,2}$ in order to continue this analysis to the $\a'^3$ order.

As far as the author is aware this is the first example of a consistent $\a'^3$ dimension-zero torsion deformation which has been given. However, it is not the full story for the heterotic string at $\a'^3$ as one also has to include the string tree-level $R^4$ contribution. From the viewpoint  of supersymmetry there is a family of such terms which could appear in the spacetime effective action which are given by integrating an arbitrary function, say $B(\f)$,  of the dilaton superfield over the full superspace, although string theory requires $B$ to be a specific function with a specific coefficient. At this order the dimension minus six auxiliary field can be non-zero and could therefore be proportional to $\a'^3 B(\f)$. One would expect the constant of proportionality to be non-zero unless there is a reason for supposing otherwise, which is certainly not obvious from a purely field-theoretic point of view. On the other hand, there might be an argument to be made from the string sigma model. In references \cite{Lechner:1987ip,Lechner:1989kk} it has been argued that it is possible to leave $T^1{}_{,2}$ unchanged and account for the tree-level $R^4$ term by amending $H_{0,3}$, roughly speaking by a term of the form $D^{11} B$. This proposal has the virtues of simplicity and economy and the authors of \cite{Lechner:1987ip,Lechner:1989kk} give some detailed calculations in support of it. Moreover, it fits in rather nicely with the fact that invariants can be obtained from $d_s$-closed $(0,3)$-forms via the ectoplasm method \cite{Berkovits:2008qw}. Whether it is correct or not will either require more detailed calculations to be carried out, or an argument to be found as to why the dimension-zero torsion should not be deformed to accommodate this effect. Let us note, however, that a non-zero $H_{0,3}$ will still give rise to an element of $H_s^3$ in the presence of a non-vanishing $T^1{}_{,2}$.

The effect of the explicit deformation of the dimension-zero torsion on the equations of motion can be obtained by systematically going through the Bianchi identities. Indeed, a general procedure for calculating the consequences of an arbitrary consistent deformation of the dimension-zero torsion has been given in $d=11$ in \cite{Cederwall:2004cg}. This will be a lengthy and tedious computation but it is necessary in order for the result to be applied to corrections to supergravity solutions. An alternative procedure would be to compute the (on-shell) invariant directly using the ectoplasm, or superform, method \cite{Gates:1997kr,Gates:1997ag,D'Auria:1982pm}, as discussed in \cite{Berkovits:2008qw}. The relevant invariant is of Chern-Simons type (the invariant will include $B_2 X_8$ in the spacetime action) and can be constructed following the usual procedure for such invariants \cite{Howe:1998tsa} starting from the exact $11$-form $W_{11}=dK_{10}$, with $W_{11}=k_7 H_3 X_8$. It was shown in \cite{Berkovits:2008qw} that this $W_{11}$ is indeed exact and that its lowest non-vanishing component, $W_{5,6}$, can be written as $t_0 K_{6,4}$; from the results of the previous section we have

\be
 K_{6,4}=ik_7 S(\c_{5,2} Y_{1,2} -\c_{1,2} Y_{5,2}) \ ,
 \la{5.1}
\ee

where $Y_{1,2}$ is the $(1,2)$-form obtained by lowering the vector index on $Y^1{}_{,2}$. The purely bosonic invariant part of the spacetime integrand will be given, roughly speaking, by evaluating four spinor derivatives acting on this expression. This gives a rather direct relation between the deformed dimension-zero torsion (proportional to $Y^1{}_{,2}$) and the higher-order correction to the spacetime action.

\vskip .8cm


{\bf Note:} This paper is based on a talk given at ``Kellyfest - a meeting in celebration of Kelly Stelle's 60th birthday'', Imperial College April 24-25, 2008.


\vskip .8cm

{\bf Acknowledgements:}  I thank Kurt Lechner and Mario Tonin for a stimulating e-mail discussion on heterotic superspace. This work was supported in part by EU grant (superstring theory) MRTN-2004-512194.

\end{document}